# How to quantify energy landscapes of solids.

Artem R. Oganov[1,2*] and Mario Valle[3]

[1] *Laboratory of Crystallography, Department of Materials ETH Zurich, HCI G 515, Wolfgang-Pauli-Str. 10, CH-8093 Zurich, Switzerland.*

[2] *Geology Department, Moscow State University, 119992 Moscow, Russia.*

[3] *Data Analysis and Visualization Services, Swiss National Supercomputing Centre (CSCS), via Cantonale, Galleria 2, 6928 Manno, Switzerland.*

\* Now at *Department of Geosciences and New York Center for Computational Science, Stony Brook University, Stony Brook 11794-2100, USA.*

**Abstract.** We explore whether the topology of energy landscapes in chemical systems obeys any rules, and what these rules are. To answer this and related questions we use several tools: (i) reduced energy surface and its density of states, (ii) descriptor of structure called fingerprint function, which can be represented as a one-dimensional function or a vector in abstract multidimensional space, (iii) definition of a „distance" between two structures, enabling quantification of energy landscapes, (iv) definition of a degree of order of a structure, (v) definitions of the quasi-entropy, quantifying structural diversity. Our approach can be used for rationalizing large databases of crystal structures and for tuning computational algorithms for structure prediction. It enables quantitative and intuitive representations of energy landscapes and reappraisal of some of the traditional chemical notions and rules. Our analysis confirms the expectations that low-energy minima are clustered in compact regions of configuration space („funnels") and chemical systems tend to have very few funnels, sometimes only one. This analysis can be applied to the physical properties of solids, opening new ways of discovering structure-property relations. We quantitatively demostrate that crystals tend to adopt one of the few simplest structures consistent with their chemistry, providing a thermodynamic justification of Pauling's 5$^{th}$ rule.

**I. Introduction.**

Energy landscapes determine the behaviour and properties of chemical systems (molecules, fluids, solids) – such as the melting temperature, strong/fragile behavious of liquids, the glass transition, dynamics and kinetics of structural transformations (e.g. displacive, order-disorder and reconstructive phase transitions, protein folding); see [1,2]. However, being a multidimensional function, an energy landscape cannot be directly visualised and even storing all values of this function on the computer would be impractical. The only way forward is to apply some reduction procedure, whereby only the most essential information is extracted and projected onto a small number of physically meaningful discriminators. What kind of information is extracted and how the discriminators are defined depends on a particular purpose, and several approaches have been proposed in the past [e.g. 3-8]. We mention in particular landscape statistics (see, e.g., [3-5]), the disconnectivity graphs [7], orientational bond order parameters [8], and various distance metrics [6,9,10]. Previous works focussed mainly on kinetics and physical properties of clusters, liquids and glasses. Our focus is on crystalline solids and simple ways of translating the immense information contained in energy landscapes into an intuitive language of chemistry.

**II. Energy surface, reduced energy surface and its density of states.**

The most important piece of information is the set of energy minima (including, but not restricted to, the global minimum). While for some purposes it is necessary also to explore the saddle points (i.e. the transition states) of the landscape, here we analyse only the

minima – i.e. metastable and stable structures, which bear relevant chemical information, can be experimentally observed and readily simulated.

The resulting reduced energy surface (i.e. the imaginary surface formed by local energy minima) is much simpler, and often has an overall funnel-like shape (Fig. 1): the further a local minimum is from the global minimum, the higher its energy. In more complex cases, there can be several energy funnels corresponding to different arrangements of chemical elements in the structure. The emergence of funnels on the energy landscape can be understood by chemical intuition, and can be traced to scale-free properties of the landscape in chemical systems [11].

The reduced energy surface is still multidimensional and overwhelmingly complex. Below we will show how to map it in one or few dimensions and to establish whether it has one or more energy funnels. Before that, we would like to dwell briefly on the useful information contained in the density of states (DOS) of the reduced energy surface – see Fig. 2.

Such DOS plots provide a wealth of important information. Typically, there is a tall maximum at high energies (denoted as "region 2" in Fig. 2), corresponding to disordered liquid-like structures, and one or several smaller maxima at lower energies separated by gaps (denoted as "region 1" in Fig. 2) and corresponding to ordered crystal structures and their defective variants[1]. The melting temperature of the crystal is related to the energy separation between the ground state and disordered structures.

As system size increases, the number of possible energy minima increases exponentially. One might think that with more degrees of freedom, larger systems would have more uniform DOS with filled energy gaps. However, just the opposite happens (Fig. 2b): the liquid-like maximum ("region 2" of the DOS) becomes higher (at the expense of ordered structures, which become more rare), narrower and more Gaussian-like (as argued in Ref. 5 using the central limit theorem), and shifted to higher energies (disorder costs energy) – see Fig. 2b. What we observe by examining energies and fingerprint functions (defined below) is that all disordered structures become nearly identical for large system sizes. In the limit of infinite system size all randomly produced structures will be identical and representative of the amorphous state, DOS will look like a delta-function, and the corresponding energy can be expected to correlate with the melting temperature. Finding the ground state with random sampling will be exponentially impossible for increasingly large systems.

---

[1] At sufficiently large system size, structures corresponding to the 2nd, 3rd, ... lowest energy minima will simply be defective versions of the lowest-energy structure.

**III. Fingerprint functions. Abstract distances, degree of structural simplicity.**

*1. Fingerprint functions.*

Here we define a function (actually, one can propose a number of similarly defined functions), which we call a fingerprint function and which uniquely characterises the structure. Such a function should be (i) derived from the structure itself, rather than its properties (such as energy), (ii) invariant with respect to shifts of the coordinate system, rotations and reflections, (iii) sensitive to different orderings of the atoms (e.g. one should be able to distinguish between different orderings in f.c.c. alloys), (iv) formally related either to experiment (diffraction patterns) or microscopic energetics, (v) robust against numerical errors. In addition, we require that fingerprints of similar structures be similar, and the difference of fingerprint functions a reasonable measure of structural (dis)similarity. In [15] we discussed definitions proposed by us [15,16] and other authors [5,9,10]. The definition introduced in [15] involves a sum over all interatomic distances $R_{ij}$:

$$\Phi(R) = \frac{N}{\sum_{cell} N_i N_j b_i b_j} \sum_{i,cell} \sum_{j} \frac{b_i b_j}{4\pi R_{ij}^2 \frac{N_j}{V} \Delta} \delta(R - R_{ij}) - 1 \quad (1)$$

The factor before the double sum is just a normalisation factor involving the total number $N$ of atoms in the unit cell, the numbers $N_i$ and $N_j$ of atoms of each type and their scalar atomic properties $b$ (which can be set, e.g., to the atomic numbers, electronegativities, chemical scale or Mendeleev numbers [13,14], or neutron scattering lengths). The double sum runs over all atoms $i$ within the unit cell and $j$ within the threshold distance $R_{max}$ of the $i$-th atom. $V$ the unit cell volume, and $\delta(R-R_{ij})$ is a Gaussian-smeared delta-function, absorbing numerical errors and making $\Phi(R)$ a smooth function. To be robust against numerical errors and computationally convenient, the $\Phi(R)$ function is discretised over bins of width $\Delta$. This $\Phi$-fingerprint is short-ranged, i.e. tends to zero at infinite $R$, and $\Phi(0) = -1$. When $b$-parameters in (1) are set to atomic neutron scattering lengths, $\Phi(R)$ is related to the structure factor:

$$S(Q) = 4\pi \int R^2 \Phi(R) \frac{\sin(QR)}{QR} dR \qquad (2)$$

When atomic numbers are used as $b$-parameters, relation (2) gives a zeroth-order approximation to the X-ray scattering structure factor.

This fingerprint, however, is not very sensitive to ordering of the atoms on a given structure and depends on the choice of $b$-parameters. To overcome these problems, we separate components of the fingerprint function coming from different atomic type pairs A-B, making the total fingerprint a matrix, each element of which is a function:

$$F_{AB}(R) = \sum_{A_i, cell} \sum_{B_j} \frac{\delta(R-R_{ij})}{4\pi R_{ij}^2 \frac{N_A N_B}{V_{cell}} \Delta} - 1 = g_{AB}(R) - 1 , \qquad (3)$$

where the double sum runs over all $i$-th atoms of type A within the unit cell and all $j$-th atoms of type B within the distance $R_{max}$. In eq. (3), $g_{AB}(R)$ is the pair correlation function; subtracting 1 from it makes it short-ranged. The $F$-fingerprint components have the same limiting behaviour as the $\Phi$-fingerprint – $F_{AB}(0) = -1$ and $F_{AB}(\infty) = 0$. Individual fingerprint components can be obtained from diffraction experiments. $F$-fingerprint does not contain atomic $b$-parameters, and is very sensitive to ordering, as well as structure.

Each *F*-component satisfies the following sum rule:

$$\int_0^\infty F_{AB}(R) R^2 dR = 0 \qquad (4)$$

*2. Distance between structures.*

Discretisation of fingerprints (1) and (3) allows them to be represented as vectors, values of the fingerprint $F_{AB}(k)$ in each *k*-th bin (of width $\Delta$) being vector coordinates. The dimensionality of the fingerprint space equals the number of bins. Each structure can thus be uniquely described by a fingerprint vector in this abstract space, and dissimilarity between structures can be calculated using one of distance definitions given below.

When there is only one atomic type, both $\Phi$- and *F*-fingerprints reduce to $g(R)-1$ (where $g(R)$ is the pair correlation function), and the Minkowski distance is:

$$D_{Minkowski} = \left[ \sum_k (F_1(k) - F_2(k))^p \right]^{1/p} \qquad (5)$$

When $p = 2$, Minkowski distances are the usual Cartesian distances. However, we found [15] that cosine distances:

$$D_{cosine} = \frac{1}{2}\left(1 - \frac{\mathbf{F_1} * \mathbf{F_2}}{|\mathbf{F_1}||\mathbf{F_2}|}\right) = \frac{1}{2}\left(1 - \frac{\sum_k F_1(k) F_2(k)}{\sqrt{\sum_k F_1^2(k)} \sqrt{\sum_k F_2^2(k)}}\right) \qquad (6a)$$

are more robust to the "curse of dimensionality" effect of distance concentration[2]. They are related to Cartesian distances in a non-linear way, but conveniently can only take values between 0 and 1 and are much more robust to small numerical errors. Cosine distances

---

[2] The high-dimensional space is almost empty and the whole concept of neighborhood is meaningless in high dimensions. Randomly generated points in highly-dimensional spaces tend to be all at the same distance (the relative contrast $(d_{max}-d_{min})/d_{min} \to 0$ as dimensionality tends to infinity) and under small perturbation the nearest point could change into the farthest one. These phenomena are not intrinsic to the high-dimensional space, but depend strongly on the distance function used. In this respect cosine distance are superior to Cartesian distance (or in general the Minkowski distances with $p > 1$)

measure the similarity between structures using the angle between their *F*-vectors. For finite systems two *F*-vectors can be parallel only if they are identical. The generalisation of (6a) to multicomponent fingerprints is straightforward:

$$D_{cosine} = \frac{1}{2}(1 - \frac{\sum_{AB}\sum_{k} F_{1,AB}(k)F_{2,AB}(k)w_{AB}}{\sqrt{\sum_{AB}\sum_{k} F_{1,AB}^2(k)w_{AB}}\sqrt{\sum_{AB}\sum_{k} F_{2,AB}^2(k)w_{AB}}}) , \qquad (6b)$$

where the sums are over all $F_{AB}$ components, taking each component (note that $F_{AB}=F_{BA}$) only once. The importance weight $w_{AB}$ of each fingerprint component is defined as:

$$w_{AB} = \frac{N_A N_B}{\sum_{cell} N_A N_B} \qquad (7)$$

Computing the distance between two structures one quantifies the difference between them. This can be used for analyzing crystal structure databases and for visualizing results of evolutionary structure prediction simulations, as shown in Fig. 4.

*3. Degree of order.*

Both $\Phi$- and *F*-fingerprints describe correlations between atomic positions– i.e. the non-randomness of the structure, responsible for its diffraction signal (eq. (2)). For an ideal gas both $\Phi$- and *F*- fingerprints are zero, and therefore their deviation from zero level can be used as a measure of order. We write a non-dimensional and scale-invariant definition of the degree of order $\Pi$:

$$\Pi = \frac{1}{\lambda}\int_0^{R_{max}} F^2(R)dR = \frac{\Delta}{\lambda}|\mathbf{F}|^2 , \qquad (8)$$

where $\lambda$ is a characteristic length (for instance, $\lambda=R_0$, the distance at which $F(R)$ first becomes zero[3]). While the angles between fingerprint vectors measure structural differences, the lengths of these vectors show the degree of order of each structure. For an ideal gas $\Pi = 0$, for a gas of hard-sphere atoms $\Pi = 1$, for solids $\Pi > 1$. We use an alternative definition of the degree of order: modifying the sum rule (4) to obtain the next non-vanishing moment we get:

$$P = \frac{1}{V/N} \int_0^{R_{max}} F^2(R) R^2 dR \qquad (9a)$$

---
[3] An alternative useful definition would be $\lambda=V^{1/3}$.

The generalisation for multicomponent fingerprints is straightforward:

$$P = \sum_{A,B} w_{AB} P_{AB} \quad , \qquad (9b)$$

where $P_{AB}$ are computed separately for each fingerprint component.

Roughly, $P$ is related to the total scattering power of the structure (*cf.* eq. (2)), and simpler structures have higher $P$ values. As we will show below on concrete cases, statistically, the higher $P$ the lower the energy of the structure and therefore the degree of order $P$ is an excellent predictor of the structural energy.

*4. Total energy and fingerprint functions.*

We can relate the total energy to *F*-fingerprint in the approximation that the total energy is a sum of pairwise interaction potentials $U_{AB}(R)$:

$$E = \frac{1}{2} \sum_{A,B} U_{AB}(R) = E_{RAN} + \frac{4\pi}{2} \sum_{A,B} \int_0^\infty F_{AB}(R) w_{AB} U_{AB}(R) R^2 dR \quad , \qquad (10)$$

where $E_{RAN}$ is the energy of a totally random structure with the same volume. Since $E_{RAN}$ is a function of volume only, energy differences *at constant volume* can be written as:

$$\Delta E_V = 2\pi \sum_{A,B} \int_0^\infty \Delta F_{AB}(R) w_{AB} U_{AB}(R) R^2 dR \qquad (11)$$

This expression shows what is intuitively expected - the more similar a structure is to the ground-state structure, the lower its energy.

*5. Quasi-entropy: a measure of structural diversity.*

Based on our distance metric, and noting that cosine distances, just like occupation numbers, take values between 0 and 1, we propose collective quasi-entropy:

$$S_{coll} = -\langle \ln(1 - D_{ij}) \rangle \tag{12}$$

as a measure of collective diversity. Calculating $S_{coll}$ for a set of structures gives a single number measuring the diversity of that set. This is very useful for tuning global optimization methods, where rapid decrease of quasi-entropy may indicate premature convergence of the algorithm.

For each given structure, one may also compute its own quasi-entropy as a measure of disorder and complexity of that structure (alternative to definitions (8) and (9)):

$$S_{str} = -\sum_A \frac{N_A}{N_{cell}} \langle \ln(1 - D_{A_i A_j}) \rangle , \tag{13}$$

where distances $D_{A_i A_j}$ are measured between fingerprints[4] of all *i*-th and *j*-th sites occupied by chemical species A, and the total quasi-entropy is a weighted sum over all chemical species. Standard crystallographic description gives the number of symmetrically inequivalent atomic positions, but does not provide any measure of physical difference between those positions – such a difference is given by eq. (13).

*6. Summary of the formalism.*

We have proposed two families of closely related fingerprint functions (1) and (3). Based on these functions, we proposed a measure (6ab) of similarity (distance) between different structures, and a new parameter (9ab) quantifying the simplicity and order of a crystal structure. Quasi-entropies (12) and (13) can be used for analysing the performance of global optimization methods and provide another measure of structural complexity. Below we show how these quantities can be used to map energy landscapes of solids and rationalise them in terms of intuitive chemical concepts.

---

[4] For this, one needs to compute a fingerprint for each *i*-th atom (rather than atomic species), defined by analogy with eq. (3) as $F_{A_i B} = \sum_{B_j} \frac{\delta(R - R_{ij})}{4\pi R_{ij}^2 \frac{N_B}{V_{cell}} \Delta} - 1$.

**IV. Illustrations of our approach on realistic systems[5].**

*1. Confirming the existence of energy funnels.*

Chemical intuition tells us that since low-energy structures share many similarities (similar coordination numbers, bond lengths and angles, sometimes even whole structural blocks), they have to be clustered in the same region of parameter space. This would lead to an overall bowl-like shape of the reduced energy surface (Fig. 1) and would have many consequences of paramount importance. First, this would imply that low-energy structures tend to be connected by relatively low energy barriers and short transition pathways (Bell-Evans-Polanyi principle [24]). If there are several funnels, transitions between them are likely to involve large activation barriers. Second, funnel-like topology of the landscape is important for modern structure prediction methods (e.g. the evolutionary algorithm [16-19] and all neighbourhood search methods) to be efficient.

If the funnel-like shape of the reduced energy landscape is valid, there would be a correlation between the energy and the distance from the global minimum: the further from the global minimum, the higher the energy. This correlation can now be easily checked and turns out to be excellent in many chemically simple systems – this is shown in Fig. 5 for GaAs and in Fig. 6 for $Au_8Pd_4$.

---

[5] In all illustrations given below, we used $F$-fingerprints (eq. (3)) with $R_{max}$=15 Å, $\Delta$ = 0.05 Å, and Gaussian-smeared δ-function with $\sigma$ = 0.075 Å. We use cosine distances and degree of order P (eq. 9a,b) throughout. The data analysed here include 2997 distinct local minima for GaAs (8 atoms/cell), 249 for $Au_8Pd_4$ (12 atoms/cell), 967 for MgO (32 atoms/cell), 949 for $H_2O$ (12 atoms/cell), 6977 for MgNH (12 atoms/cell), and 1949 distinct local minima for the $AB_2$ Lennard-Jones crystal. All data were obtained in evolutionary and random sampling runs, duplicate minima were pruned using a clustering procedure described in [15]. Local optimizations for most systems were done within the generalized gradient approximation [20] (only for $Au_8Pd_4$, calculations were done within the local density approximation) using the VASP code [21]. Exceptions are MgO, for which we used the interatomic potential [22] and Lennard-Jones (see caption to Fig. 2 for the description of the potential model) systems – for both, structure relaxations and energy calculations were performed using the GULP code [23].

At first surprisingly, for MgO with 32 atoms/cell the landscape turns out to be more complex and contains three broad funnels with 6-, 5- and 4-coordinate Mg atoms, respectively (Fig. 7a). The potential [22] used here correctly finds the ground state to have the NaCl-type structure, and among the lowest-energy local minima we find an exciting metastable structure with both Mg and O atoms in the fivefold coordination (Fig. 7b). This structure was previously predicted in [26], but has not yet been identified experimentally. Among other low-energy metastable structures are zincblende, wurtzite and their polytypes. The ability of Mg atoms to adopt very different coordination numbers (ranging from 4 to 8) in oxides and silicates is well known to mineralogists, and is the key to the multi-funnel structure of the landscape. Fig. 7a shows that the funnels almost "touch" each other and there are low-energy structures close to the boundaries between funnels. This means that there may be relatively low-energy transition pathways between the funnels, and the landscape can be considered intermediate between single-funnel and multi-funnel types. The non-monotonic energy-distance plot reflects the complicated topology of the landscape very clearly (Fig. 7c).

An extremely clear depiction of a multi-funnel energy landscape is provided by $H_2O$ with 12 atoms/cell. This landscape contains (at least) three clearly separated and chemically distinct funnels (Fig. 8). These funnels have very different energies – structures in the deepest funnel consist of $H_2O$ molecules, whereas structures belonging to the two higher-energy funnels contain various mixtures of molecular groups ($H_2O$, $H_2$, $O_2$, $OH^-$, many structures contain $H_2O_2$ and some $H_2O_3$ molecules). In many structures we find $OH^-$ ions, which is at variance with previous calculations [27] where these ions were not seen at all.

The energy-distance correlation is clearly seen across the landscape and, of course, within each funnel (Fig. 8b).

Among the systems we explored, the most complex landscapes are found for MgNH and $AB_2$ Lennard-Jones crystal with the potential described in caption to Fig. 2. Both systems, though small (12 atoms/cell), possess remarkable complexity with multiple degeneracies.

For instance, for MgNH we found 25 structures with energies less than 2.5 meV/atom above the ground state. Most of these structures contain $NH^{2-}$ groups, though some have $(NH_2)^-$ and $N^{3-}$ ions - low energies of the latter structures suggest (in agreement with experiment) that MgNH is only marginally stable with respect to decomposition into $Mg_3N_2+Mg(NH_2)_2$. Generally, the existence of funnels with energies close to the ground state indicates that the system is close either to a phase transition or to decomposition. The 2D-map of the landscape and the energy-distance correlation plot for MgNH (Fig. 9a) show low-energy funnels, some of which show very obvious structural differences. One of the funnels consists entirely of layered structures, whereas in another funnel all structures contain either parallel or antiparallel $NH^-$ groups. Finally, we note the gap in the energy distribution, related to the dissociation of molecular ions – the higher-energy structures contain atomic/ionic hydrogen, the formation of which requires breaking N-H bonds.

For the $AB_2$ Lennard-Jones crystal the energy-distance plot (Fig. 9b) has several flat regions with nearly constant energies, a feature characteristic of strong degeneracy. We will see in the next section that several families of structures exist here, and within each family the energy is nearly constant. The degeneracies originate from the short-rangedness of interatomic interactions, where the length scale of typical geometric features (in this case, interplanar distances) is longer than the range of interatomic interactions, and in the competition between different interactions (in this model, A-A, B-B and A-B interactions have the same strength and thus compete). Such degeneracies and complex energy landscapes (probably less extreme than the case presented in Fig. 9b) may be expected in systems forming quasicrystals and incommensurate phases, as well as systems unstable (as the $AB_2$ crystal here) or nearly unstable (as MgNH) agaist decomposition.

*2. Does nature prefer simple structures?*

Pauling's fifth rule [28] states that "*the number of essentially different kinds of constituents in a crystal tends to be small*" – in other words, structures tend to be simple. Some have accepted this rule, some criticized it. Our approach enables its systematic analysis from the viewpoint of structural energies. If Pauling's rule is valid, one should see a clear correlation between the energy and degree of order. Finding cases where the correlation breaks down would indicate the limits of its applicability.

A vast majority of cases that we analysed shows an excellent correlation between the energy and the degree of order. This correlation is equally good for single- and multi-funneled landscapes; in other words, it is more fundamental. For GaAs, MgO and $H_2O$ ground states are *the* most ordered structures among the multitude of structures that we generated. For $Au_8Pd_4$ the ground state is among very few most ordered structures, and even for heavily frustrated systems, such as MgNH and $AB_2$ Lennard-Jones crystal, the ground states still belong to a small number of most ordered structures. Fig. 10 shows such correlation plots for MgO, $Au_8Pd_4$, $H_2O$ and Lennard-Jones $AB_2$ crystal.

Furthermore, we see that all chemically interesting structures are on the high-order side of the correlation plots. We illustrate this point for MgO (Fig. 10a), where a number of interesting and even unexpected low-energy ordered structures were found with the help of such a correlation plot. For $AB_2$ Lennard-Jones crystal, we again see families of structures with nearly identical energies. The two most ordered families include a family of defective NaCl-type structures (cubic close-packings of the B atoms, with a half of octahedral voids occupies by the A atoms – the well-known $CdCl_2$ structure type also belongs to this family) and a family of layered structures, where the B atoms form a defective hexagonal close packing, interrupted by layers of the A atoms in the trigonal prismatic coordination (Fig. 11). It is the latter family that contains several degenerate ground-state structures. The A-layers are topologically identical to the graphene sheet, and there is a striking structural similarity between the ground states of the $AB_2$ Lennard-Jones crystal and Al-C alloys [29]. Another similarity is that both Al-C alloys and the $AB_2$ Lennard-Jones crystal are unstable to decomposition into pure elements.

Recently, Hart [30] found that for the special case of ordering in alloys the simplest ordering schemes have the highest probability of occurring[6]. Here we have generalised this conclusion to all structures (i.e. beyond the special case of ordering in alloys) and confirmed it by analysing structural energies. We can conclude that: *The ground state normally adopts one of the simplest structures compatible with the chemistry of the compound. Such structures tend to have lower energies.* This rule is nothing else than an energetics-based reformulation of Pauling's fifth rule. The main limitation of this rule is the presence of competing interactions, as in the case of $AB_2$ Lennard-Jones crystal (Fig. 10 e). Structural quasi-entropy $S_{str}$, which is zero when each atomic species occupies only one Wyckoff orbit and increases as the physical difference between the occupied sites increases, is another and perhaps even more direct measure of structural simplicity. It more directly relates to the original [28] formulation of Pauling's 5$^{th}$ rule. Unlike the degree of order, $S_{str}$ does not depend critically on the smoothing function $\delta(R-R_{ij})$ in the fingerprint definition (eqs. 1,3) and its absolute (not only relative) values are meaningful. For all cases investigated here, we found an excellent correlation between the energy and quasi-entropy $S_{str}$ – this is shown in Fig. 12 for MgO (32 atoms/cell) and MgNH (12 atoms/cell). Like with the degree of order *P*, the correlation is worst for the AB$_2$ Lennard-Jones crystal, the reason being its instability to decomposition that leads to long-period layered structures that contain quite diverse atomic sites. The same can be expected for other complex and frustrated systems. A good example of which is the recently discovered [31] high-pressure stable phase of boron, the structure of which contains atomic sites that are so different that there is charge transfer between them [31] and the structure possesses large $S_{str}$=0.18. which formulates an intuitively reasonable expectation that in stable structures atoms occupy. Our conclusion is that: *In the ground state and low-energy structures, atoms of each species tend to occupy similar crystallographic sites.*

---

[6] To quantify the simplicity of an alloy structure, Hart [30] introduced an index measuring the non-randomness of the distribution of atoms over sites. His index is valid only for the case of ordering in alloys. Our degree of order (eq. 9a,b) provides a universal criterion, equally valid for ordering in alloys of a given structure, as well as for the comparison of completely different structures.

The degree of order *P* and structural quasi-entropy $S_{str}$ are not the only possible measures of structural simplicity. One alternative is the set of orientational bond order parameters $Q_n$ proposed by Steinhardt [8]. These parameters successfully differentiate between liquid-like and crystal-like configurations and could be expected to be good predictors of the energy. However, for all systems examined here (even as simple as MgO) the correlation of these parameters with the energy is either very weak or non-existent – Fig. 13 shows this for the $Q_6$ parameter (we also checked for $Q_4$, as well as $W_4$ and $W_6$ parameters [3132], and arrived at the same conclusion). Thus, the degree of order *P* and structural quasi-entropy $S_{str}$ are strong predictors of the energy, whereas the orientational bond order parameters $Q_n$ and $W_n$ are not.

**Conclusions.**

We have introduced a number of powerful tools to investigate structures and energy landscapes of crystalline systems. These tools can be used for rationalising structural and thermodynamic information on solids.

The basic function, from which all others are derived, is the fingerprint (eq. 3), which can be represented as a vector in an abstract multidimensional space. The possibility of computing a well-defined index of similarity ("distance") between two crystal structures is valuable for many purposes – as two examples, we mention monitoring of the progress of structure prediction simulations [16] and the emerging field of crystallographic genomics [33], where such analysis is central for the rationalisation of large databases of crystal structures.

Typical dimensionalities of fingerprint vector spaces for cases studied here, $10^2$-$10^3$, are clearly redundant compared to the true dimensionality $d = 3N + 3$ ($N$ is the number of atoms in the unit cell). However, even $d$ overestimates the true dimensionality of the landscape, because it ignores short-range order that leads to certain constraints $\kappa$ in relative positions of the atoms. The actual intrinsic dimensionality:

$$d^* = 3N + 3 - \kappa \qquad (14)$$

is generally a non-integer number and can be computed from the distance distribution. Using the Grassberger-Procaccia algorithm in the modification [34], we obtain $d^* = 10.85$ ($d = 39$) for $Au_8Pd_4$, $d^* = 32.5$ ($d = 39$) for MgNH and $d^* = 11.6$ ($d = 99$) for MgO. The intrinsic dimensionality gives the minimum number of dimensions sufficient for mapping the data exactly, but approximate mappings can be done in lower dimensions.

By allowing direct mapping of energy landscapes in any number of dimensions – from 1 (energy-distance plots extensively used here) to 2 (landscape "maps" also presented here) to as many dimensions as needed, such analysis leads to a new level of chemical insight.

Based on the fingerprint function, we introduced two indices of structural simplicity, the degree of order $P$ (eq. 9ab) and structural quasi-entropy $S_{str}$ (eq. 13), which enabled us to directly verify the well-known Pauling's fifth rule (the parsimony rule): statistically, simpler structures tend to have lower energies. Parameters $P$ and $S_{str}$ are powerful tools for analyzing results of crystal structure prediction simulations (see [15]) and crystal structure databases. This analysis can be brought further: for instance, energy-order correlations for order components $P_{AB}$ (eq. 9b) may indicate the dominant structure-forming interactions and causes of geometric frustrations.

The tools developed here are helpful in monitoring structure prediction simulations, as shown in Fig. 14. Furthermore, these tools, as well as improved understanding of energy landscapes, can be (and already have been [35]) used to improve performance of the existing global optimization methods for crystal structure prediction.

The same analysis can be done for the physical properties of crystals, leading to the discovery of interesting structure-property relations. For many properties we also expect to observe funnel-shapes landscapes, which would indicate types of structures that are of particular interest. Many properties (e.g. the electrical conductivity) are expected to be well correlated with the degree of order $P$ and structural quasi-entropy $S_{str}$. The analysis presented here can thus provide the general link between the structure, properties and stability of solids.

**Acknowledgments.** The formalism discussed here has been implemented by M.V. in the STM4 toolkit, freely available toolkit, freely available at: http://www.cscs.ch/~mvalle/STM4; fingerprint libraries can be found at: http://trac.cscs.ch/crystalfp. We greatly appreciate discussions with A.O. Lyakhov. This work was supported by Swiss National Science Foundation (grants 200021-111847/1 and 200021-116219).

**FIGURE CAPTIONS:**

**Fig. 1. Conceptual depiction of an energy landscape (solid line) and reduced energy landscape (filled squares connected by dashed lines).** From [16].

**Fig. 2. DOS of energy minima for (a) GaAs with 8 atoms/cell and (b) binary Lennard-Jones crystal $AB_2$ for several system sizes.** Energies are shown relative to the ground state. Data were obtained by locally optimising random structures, without pruning identical structures. In (a), 3000 random structures were sampled, while in (b) we typically sampled 5000 random structures at each system size. Calculation (a) was done using the generalized gradient approximation [20] of density functional theory. In (b), for each atomic pair the Lennard-Jones potential was written as $U_{ij} = \varepsilon_{ij}\left[(\frac{R_{min,ij}}{R})^{12} - 2(\frac{R_{min,ij}}{R})^6\right]$, where $R_{min,ij}$ is the distance at which the potential reaches minimum, and $\varepsilon$ is the depth of the minimum). For all pairs we used the same $\varepsilon$, but different ideal lengths: $R_{min,BB}=1.5R_{min,AB}=2R_{min,AA}$. Competition between these simple spherically-symmetric interactions leads to very complex energy landscapes and non-trivial ground states.

**Fig. 3.** *F*-**fingerprint of the ground-state structure of $Au_8Pd_4$ (12 atoms/cell) found in [16].**

**Fig. 4. Similarity matrix (dimensions 630x630) for an evolutionary structure prediction simulation for $Au_8Pd_4$ (with 12 atoms/cell) at 1 atm.** Each (N,M) pixel shows the distance between the N-th and M-th structures. Note the increase of similarity between the structures towards the end of the simulation – this is a consequence of "learning" in evolutionary simulations. This matrix is by construction symmetric with respect to its diagonal, and all diagonal (N,N) elements are zero (structure N is identical to itself).

**Fig. 5. Energy-distance correlation for 2997 distinct local minima of GaAs (8 atoms/cell). The minima were found in several *ab initio* evolutionary and random sampling simulations.** Energy differences (per unit cell) and distances in this and all subsequent graphs are relative to the ground state.

**Fig. 6. Energy-distance correlation based on 249 distinct local minima of $Au_8Pd_4$ obtained in an *ab initio* evolutionary run [24].** Most structures are different decorations of *f.c.c.* or *h.c.p.* structures.

**Fig. 7. Energy landscape for MgO with 32 atoms/cell.** (a) 2D-mapping of the landscape, where each point represents a structure and distances between points on the graph are maximally close to the distances between the corresponding fingerprints[7]. Darker points indicate lower-energy structures. (b) Metastable structure of MgO with 5-coordinate Mg and O atoms in the trigonal bipyramidal coordination. It has the space group $P6_3/mmc$, cell parameters $a=b$=3.46 Å and $c$=4.18 Å, and atomic positions Mg (1/3, 2/3, 3/4) and O (2/3, 1/3, 3/4). (c) Energy-distance correlation.

**Fig. 8. Energy landscape of $H_2O$ with 12 atoms/cell.** (a) 2D-map of the landscape with darker points indicating lower-energy structures. (b) Energy-distance correlation. The inset shows only the 580 structures based on $H_2O$ molecules.

---

[7] Of course, such mapping with reduced dimensionality cannot fully preserve distances (for the same reason, 2D-maps of the world show distorted distances). However, the topology of the landscape is correctly reproduced.

**Fig. 9. Complex energy landscapes :** (a) MgNH (12 atoms/cell) and (b) $AB_2$ Lennard-Jones crystal (12 atoms/cell).

**Fig. 10. Energy-order correlation plots.** (a) MgO (32 atoms/cell), (b) $Au_8Pd_4$ (12 atoms/cell), (c) $H_2O$ (12 atoms/cell), with inset showing only the 580 structures containing only $H_2O$ molecules, (d) MgNH (12 atoms/cell) and (2) $AB_2$ Lennard-Jones crystal (12 atoms/cell). Large red squares indicate the ground state.

**Fig. 11. Families of nearly degenerate structures for $AB_2$ Lennard-Jones crystal (12 atoms/cell).** (a-c) ground states with layers of the A atoms in the trigonal prismatic coordination, (d,e) defective NaCl-type structures. Structure (a) is the lowest-energy structure we found.

**Fig. 12. Energy – structural quasi-entropy $S_{str}$ correlation plots. (a) MgO (32 atoms/cell), (b) MgNH (12 atoms/cell).** Note a clear energetic preference for simple structures (those with lowest quasi-entropy).

**Fig. 13. Energy-$Q_6$ correlation plots. (a) MgO (32 atoms/cell), (b) $AB_2$ Lennard-Jones crystal (12 atoms/cell).** It is obvious that, unlike the degree of order $P$ (eq. 9ab) or structural quasi-entropy (eq. 13), the orientational bond order parameter $Q_6$ is not a good predictor of energy.

**Fig. 14. Progress of an evolutionary structure prediction simulation of the $AB_2$ Lennard-Jones crystal (12 atoms/cell).** Despite complexities of the energy landscape of this system (the ground state was found only once after relaxing 5000 random structures), the evolutionary algorithm USPEX [1619] finds the ground state after only 110 structure

relaxations. The initial population consisted of 50 random structures, and each subsequent generation consisted of 10 best old non-identical structures and 10 new structures produced from them using the variation operators described in [1619]. In each generation the collective quasi-entropy $S_{coll}$ was computed for a set of structures that were used for producing the next generation. $S_{coll}$ should not decrease too fast, and when it becomes sufficiently small, the simulation can be terminated. *Inset* shows the lowest energy as a function of generation number.

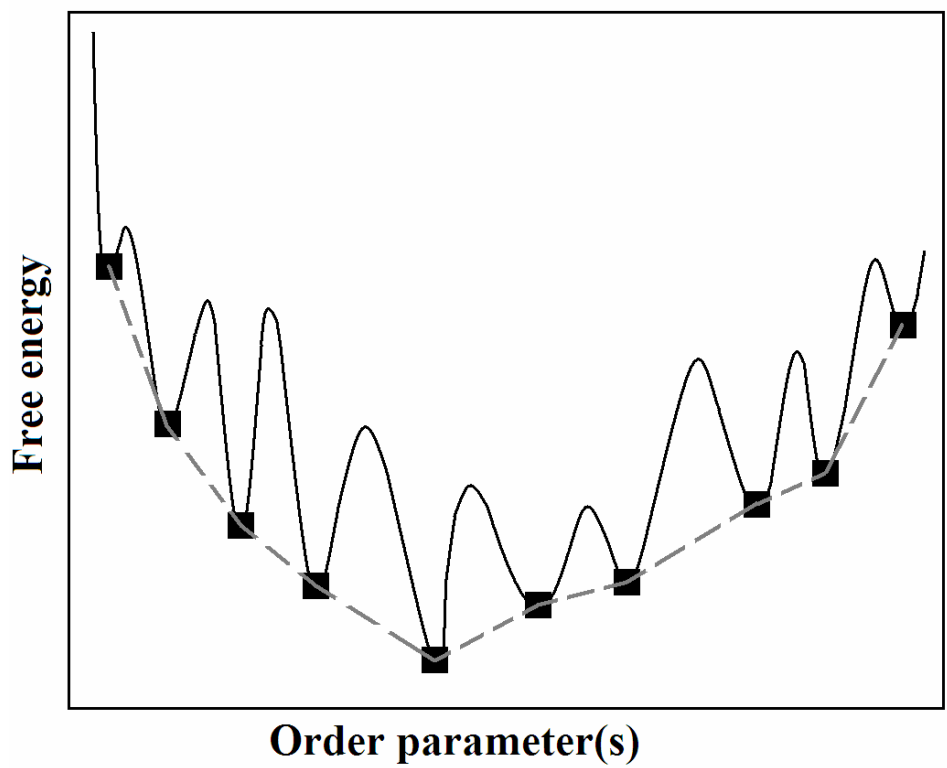

**Fig. 1.**

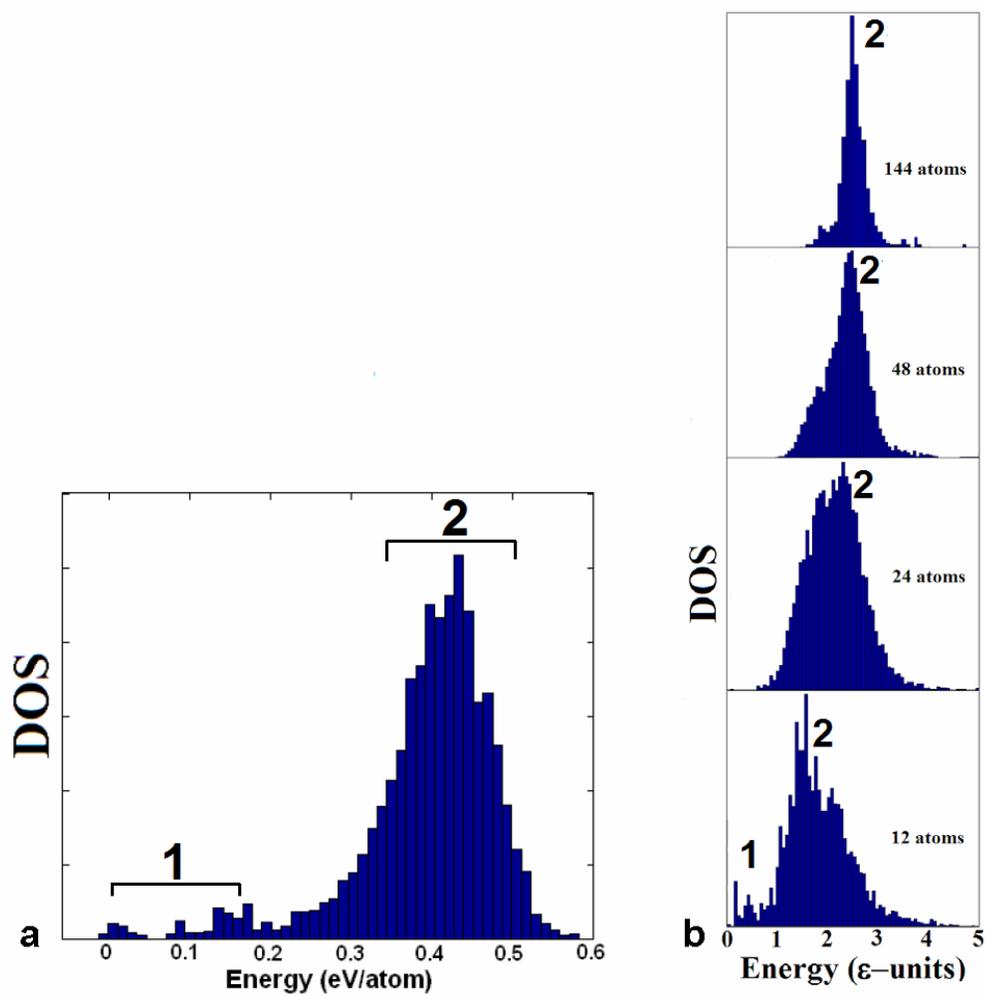

**Fig. 2.**

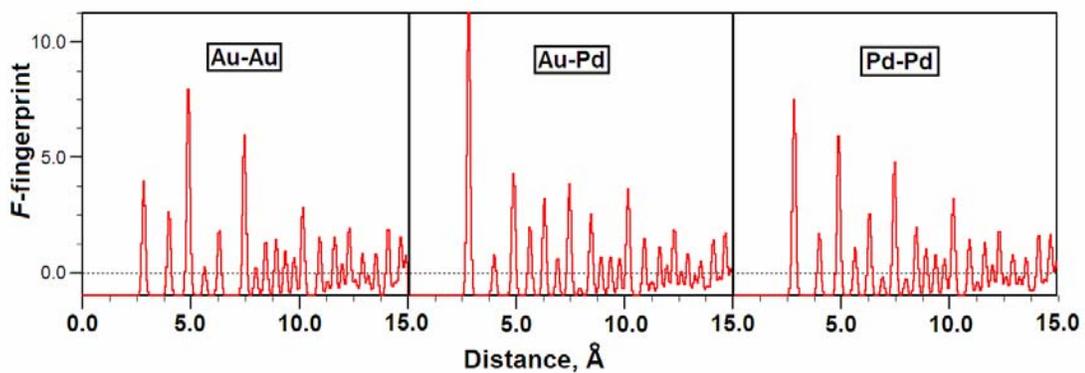

**Fig. 3.**

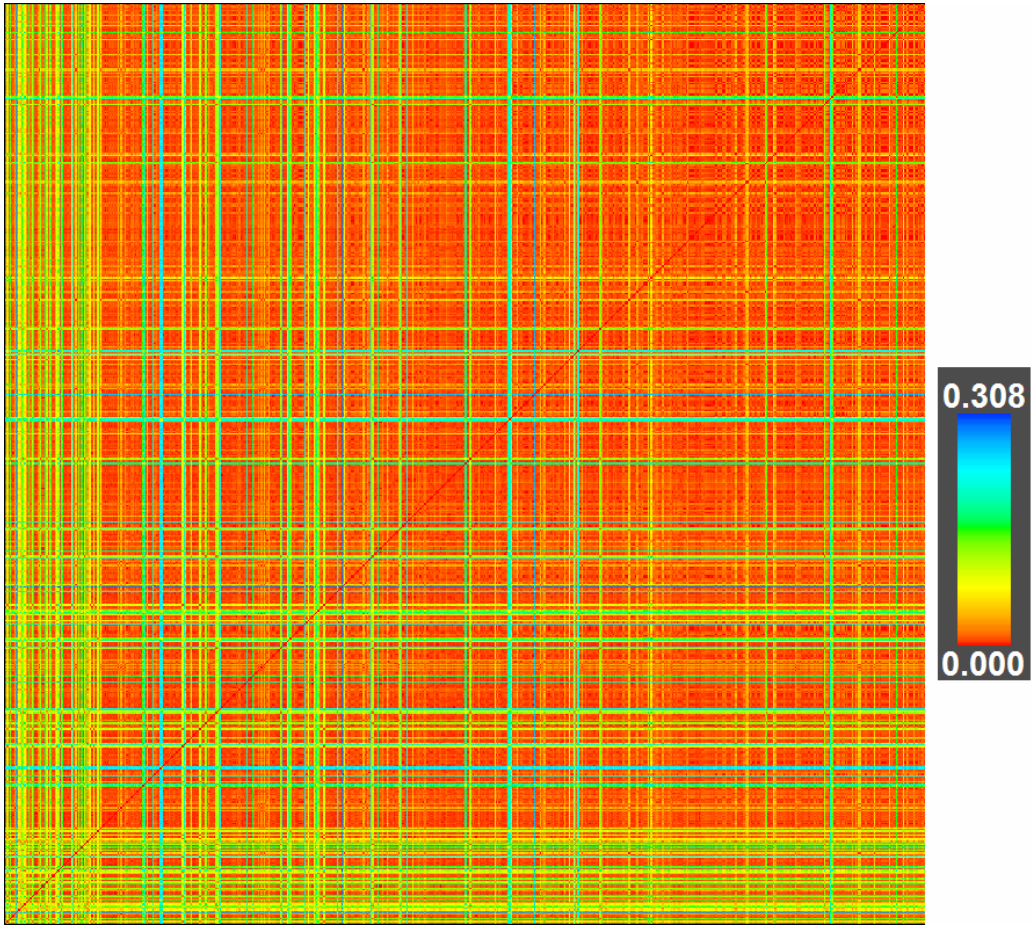

**Fig. 4.**

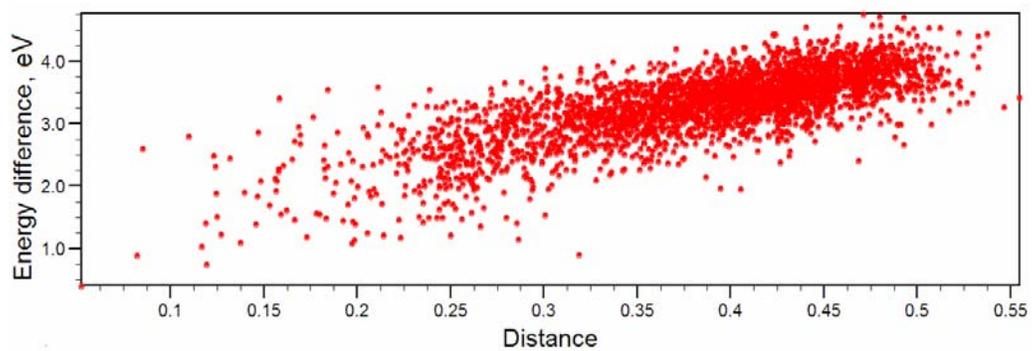

**Fig. 5.**

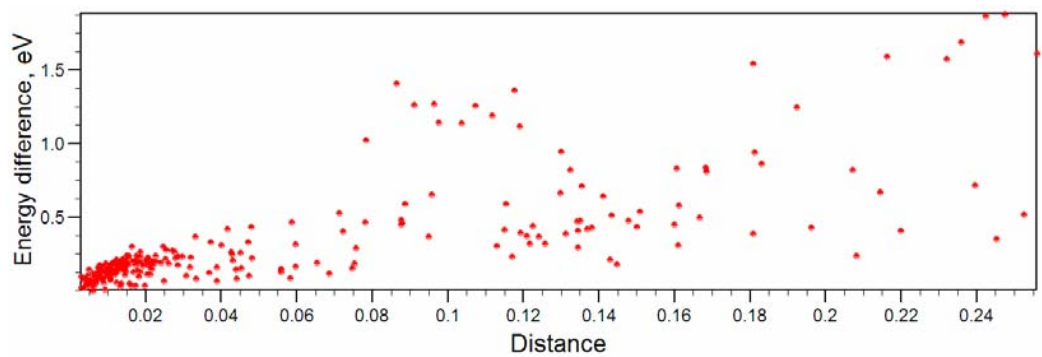

**Fig. 6.**

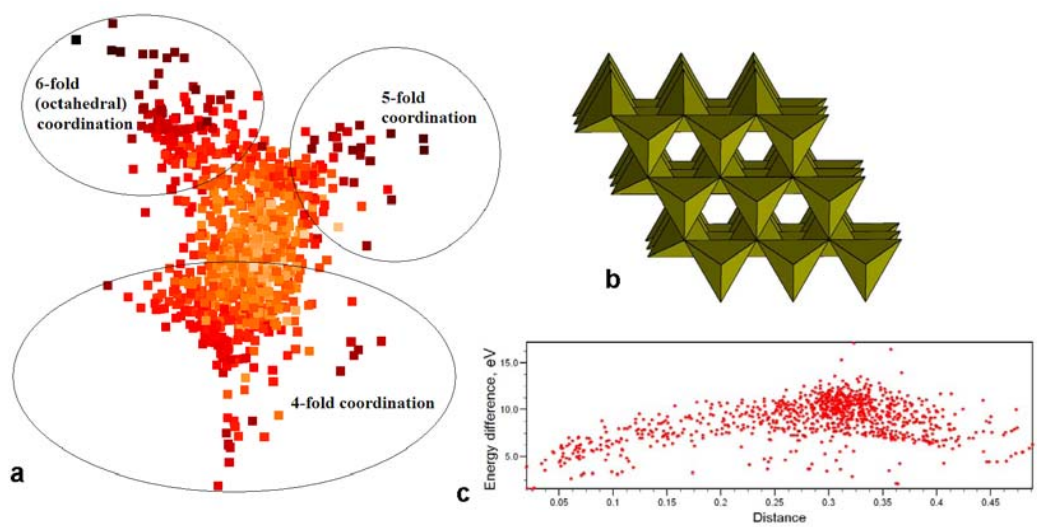

**Fig. 7.**

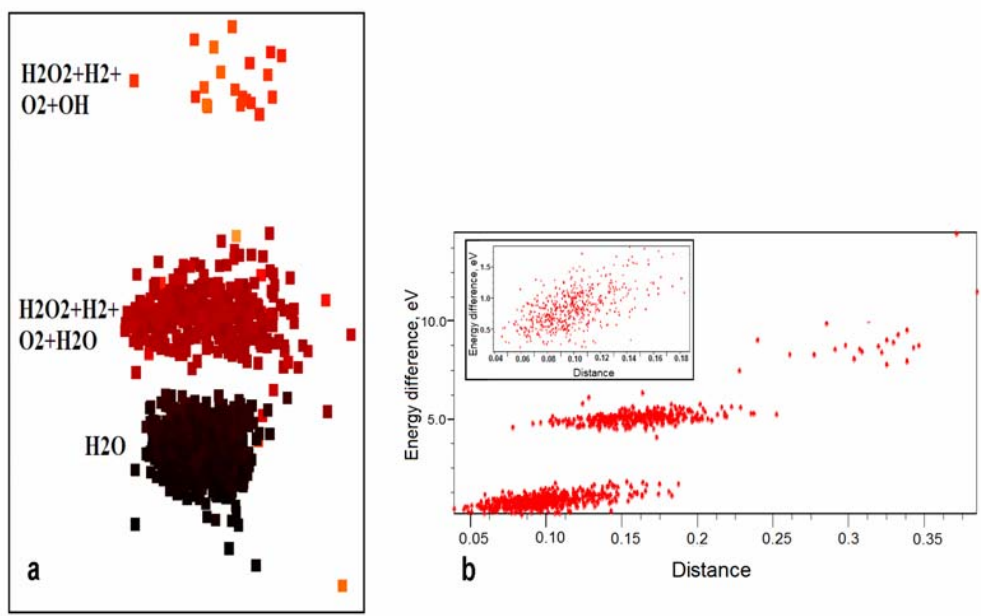

**Fig. 8.**

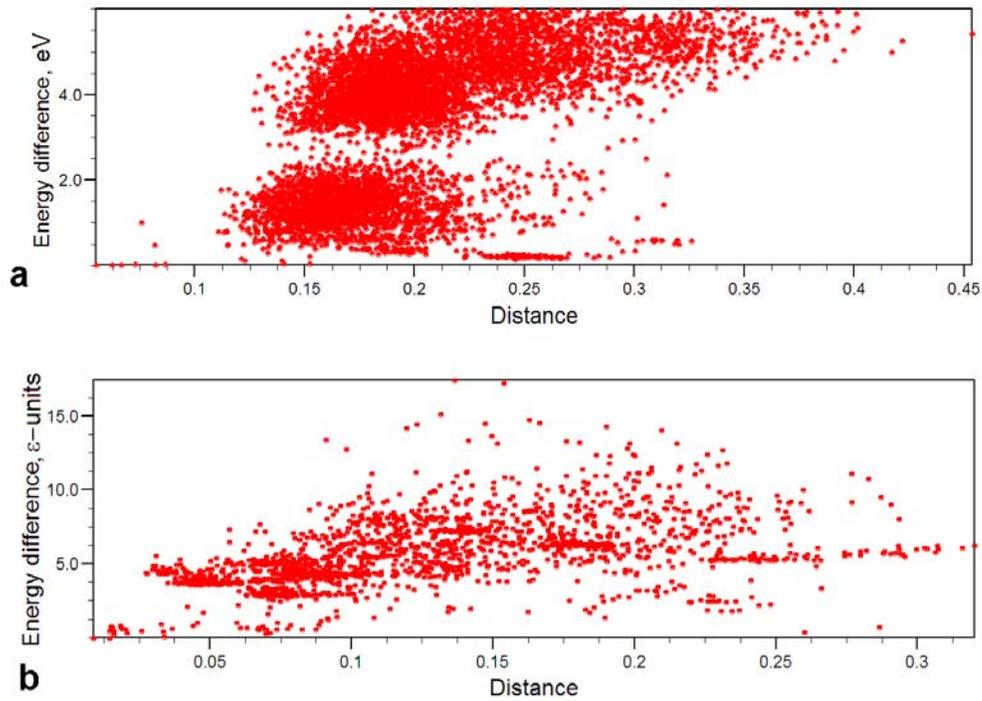

**Fig. 9.**

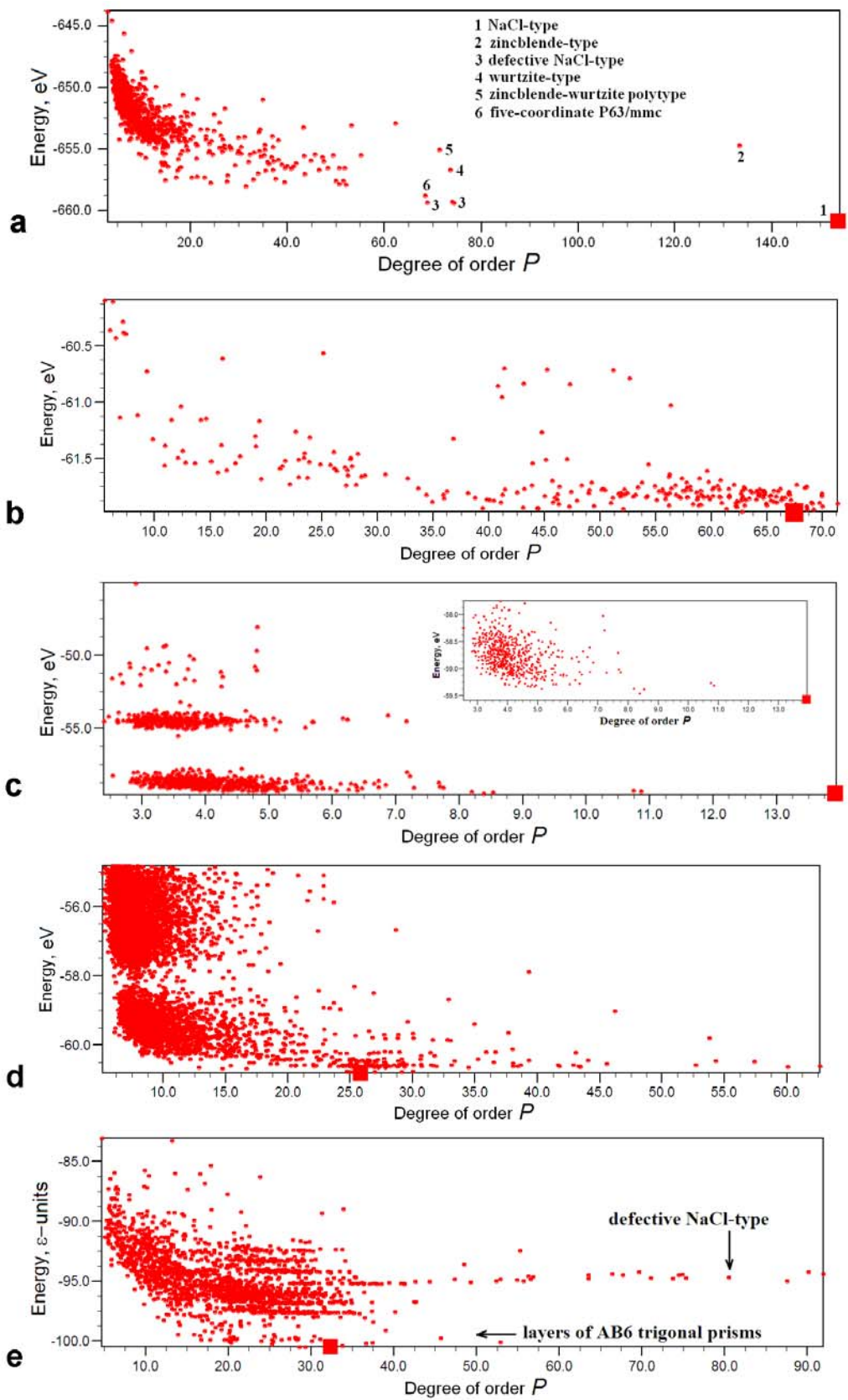

**Fig. 10.**

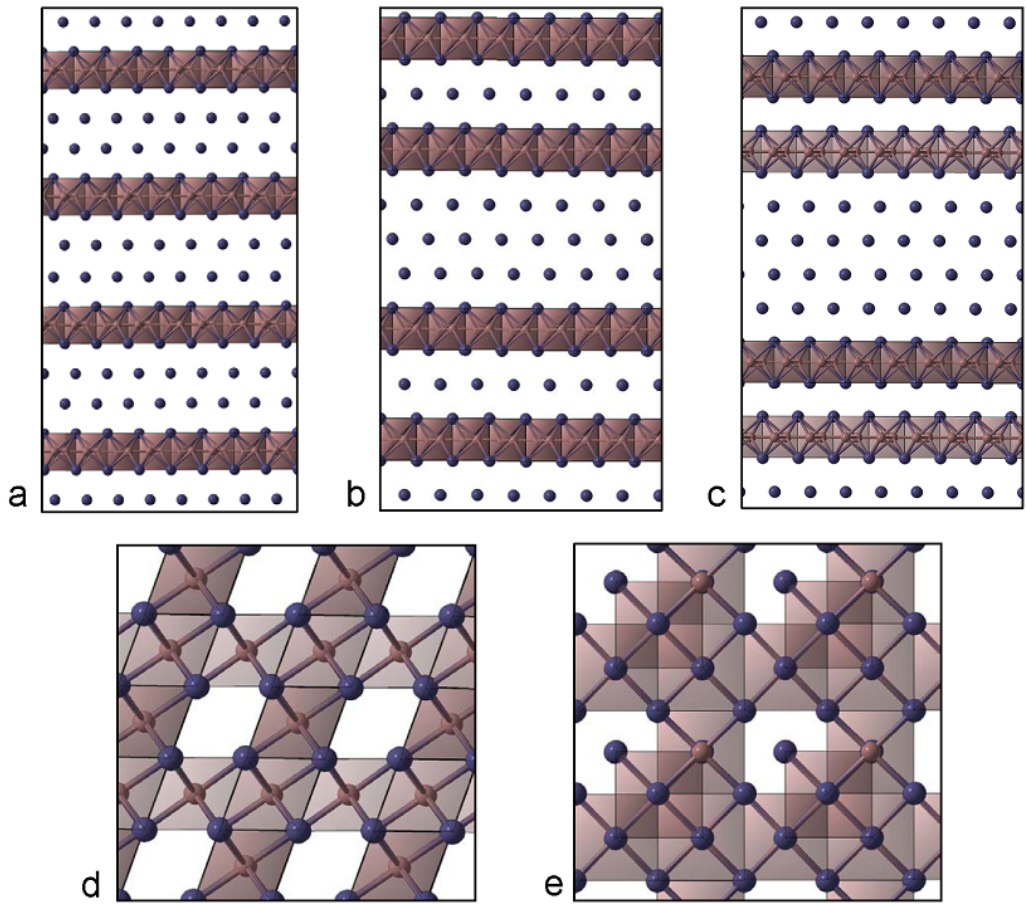

**Fig. 11.**

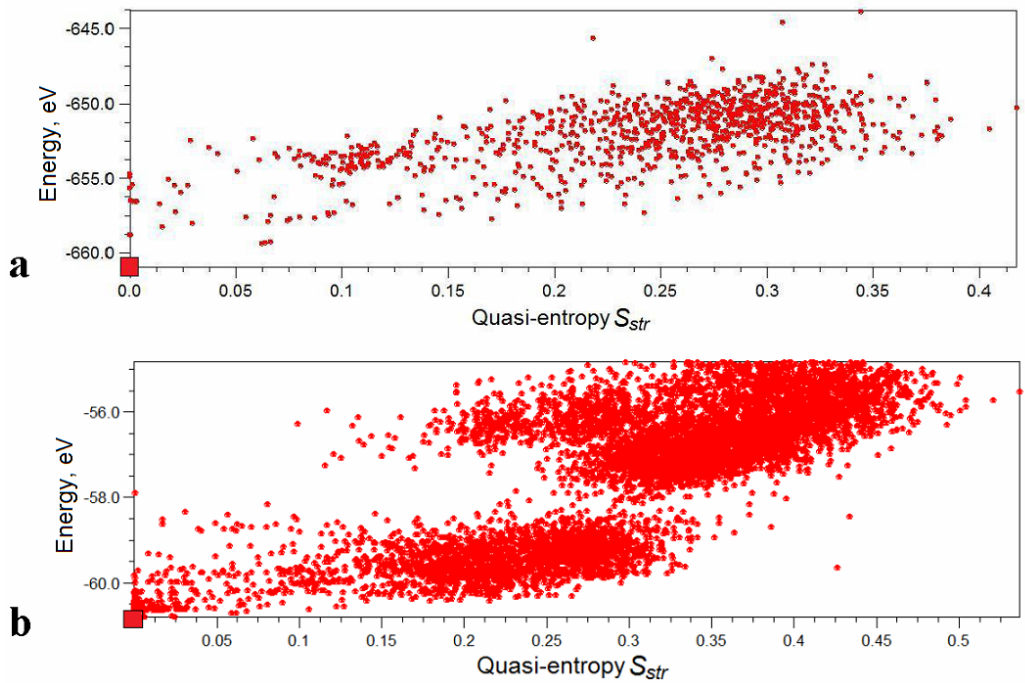

**Fig. 12.**

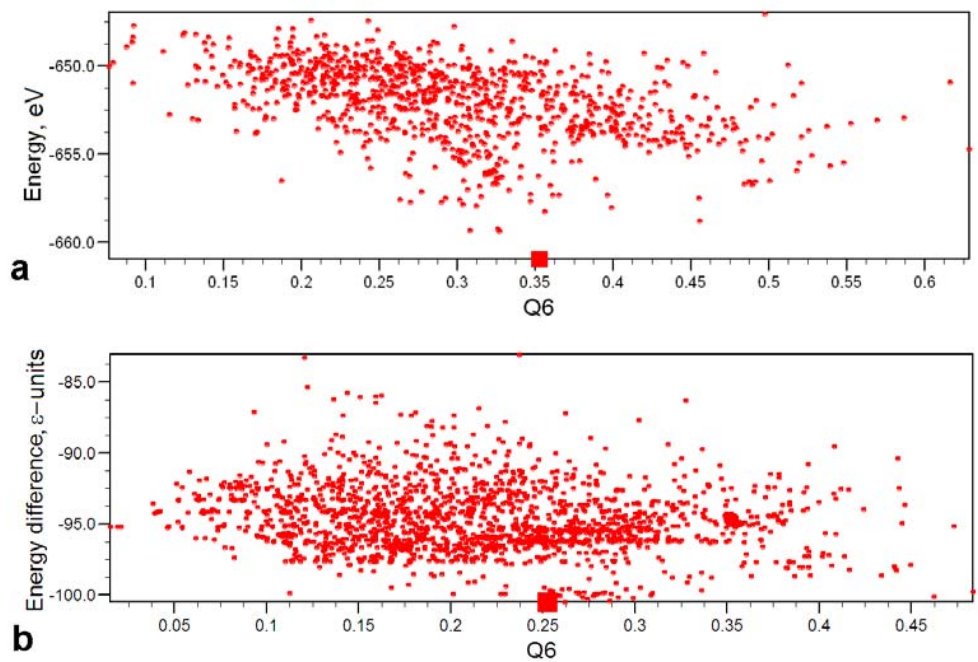

**Fig. 13.**

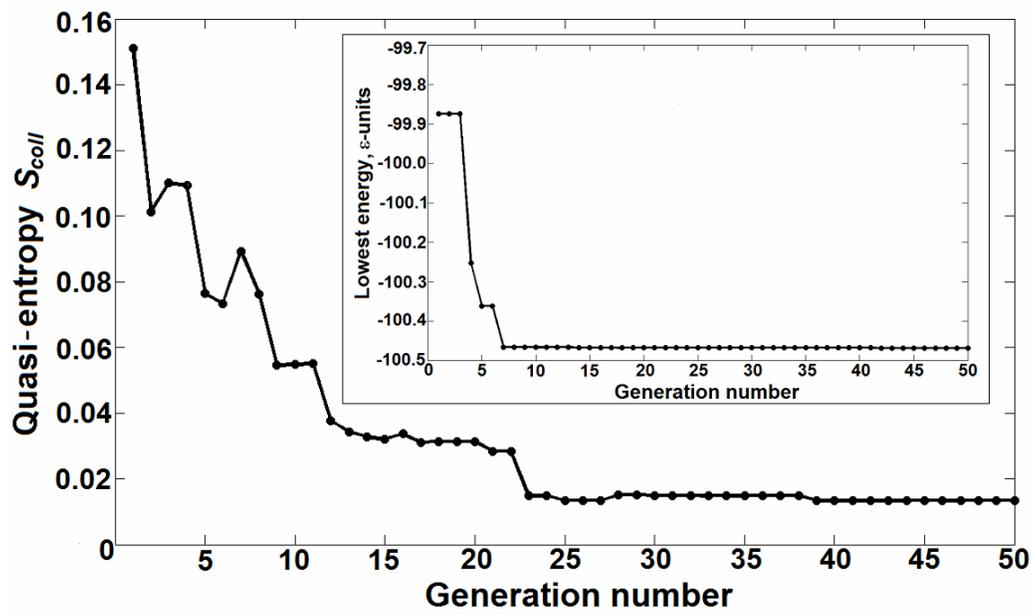

**Fig. 14.**